\documentclass[english]{revtex4-2}
\usepackage[T1]{fontenc}
\usepackage[latin9]{inputenc}
\usepackage{color}
\usepackage{babel}
\usepackage{mathrsfs}
\usepackage{amsbsy}
\usepackage{amstext}
\usepackage{amssymb}
\usepackage{esint}
\usepackage[unicode=true,pdfusetitle,
 bookmarks=true,bookmarksnumbered=false,bookmarksopen=false,
 breaklinks=false,pdfborder={0 0 1},backref=false,colorlinks=true]
 {hyperref}
\begin{document}
\title{Gauging the Maxwell extended $\mathcal{GL}\left(n,\mathbb{R}\right)$
and $\mathcal{SL}\left(n+1,\mathbb{R}\right)$ algebras}
\author{Salih Kibaro\u{g}lu$^{1,2}$}
\email{salihkibaroglu@maltepe.edu.tr}

\author{Oktay Cebecio\u{g}lu$^{3}$}
\email{ocebecioglu@kocaeli.edu.tr}

\author{Ahmet Saban$^{3}$}
\email{ahmetsaban55@gmail.com}

\date{\today}
\begin{abstract}
We consider the extension of the general-linear and special-linear
algebras by employing the Maxwell symmetry in $D$ space-time dimensions.
We show how various Maxwell extensions of the ordinary space-time
algebras can be obtained by a suitable contraction of generalized
algebras. The extended Lie algebras could be useful in the construction
of generalized gravity theories and the objects that couple to them.
We also consider the gravitational dynamics of these algebras in the
framework of the gauge theories of gravity. By adopting the symmetry-breaking
mechanism of the Stelle--West model, we present some modified gravity
models that contain the generalized cosmological constant term in
four dimensions.
\end{abstract}
\affiliation{$^{1}$Department of Basic Sciences, Faculty of Engineering and Natural
Sciences, Maltepe University, 34857, Istanbul, Turkey}
\affiliation{$^{2}$Institute of Space Sciences (IEEC-CSIC) C. Can Magrans s/n,
08193 Barcelona, Spain}
\affiliation{$^{3}$Department of Physics, Kocaeli University, 41380 Kocaeli, Turkey}
\maketitle

\section{Introduction}

The historical developments show that the concept of gauge symmetry
is a very powerful principle for constructing theories of fundamental
interactions. For~example, the~electro-weak and strong interactions
are described by the gauge theory based on the internal symmetry groups
$SU\left(2\right)\otimes U\left(1\right)$ and $SU\left(3\right)$,
respectively. In~addition to these treatments, inspired by the Yang--Mills
gauge theory~\citep{Yang:1954Conservation}, Einstein's general theory
of relativity can be considered as a gauge theory of gravity. In~1956,
Utiyama proposed that gravity can be constructed as a gauge theory
based on the local homogeneous Lorentz group~\citep{Utiyama:1956Invariant}.
Later, in~1961, Kibble and Sciama generalized the gauge group to
the {Poincare} 
group and they arrived at what is now known as the Einstein--Cartan
gravity~\citep{Kibble:1961LorentzInvariance,Sciama1962analogy}.
After~that, many space-time symmetry groups have been used to construct
different types of gauge theories of gravity, such as Weyl~\citep{Bregman:1973Weyl,Charap:1973GaugeWeyl},
affine~\citep{Borisov:1974AffineConformal,Hehl:1978HyperMomentum,Lord:1978MetricAffine}
and conformal groups~\citep{CrispimRomao:1977SuperConformal,Kaku:1977Conformal,Kaku:1977Superconformal}.

The invariance of a given system under a certain symmetry transformation
helps to find its physical properties. From this idea, if we use new
symmetries or extend the well-known symmetries (such as the Poincare
group, the de-Sitter group, etc.) it is expected that one may get
more information about any physical systems. Thus, one can say that
new or enlarged symmetries may have a great potential to formulate
any physical system more accurately. In this context, the~Maxwell
symmetry introduced by~\citep{bacry1970group,schrader1972maxwell}
is a good example of the symmetry extension, wherein the Poincare
group which describes the symmetries of empty Minkowski space-time
is enlarged by six additional abelian anti-symmetric tensor generators
satisfying 
\begin{eqnarray}
\left[P_{a},P_{b}\right] & = & iZ_{ab}.
\end{eqnarray}
This enlarged symmetry naturally extends the space-time geometry.
Physically, this extension can be considered as the symmetries of
a charged particle in a constant electromagnetic field background
\citep{bonanos2009infinite}.

Moreover, the~Maxwell symmetry has attracted increasing attention
after the work of Soroka~\citep{soroka2005tensor}. Since then a
variety of different Maxwell (super) symmetry algebras with interesting
geometric and physical properties have been constructed and analyzed.
For~example, in general, the~gauge theory of gravity based on the
extended algebras leads to a generalized theory of gravity that includes
an additional term to the energy-momentum tensor together with the
cosmological constant~\citep{azcarraga2011generalized,durka2011gauged,soroka2012gauge,azcarraga2014minimalsuper,cebeciouglu2014gauge,cebeciouglu2015maxwell,concha2015generalized,kibarouglu2019maxwellSpecial,kibarouglu2019super,kibarouglu2020generalizedConformal,kibarouglu2021gaugeAdS,cebeciouglu2021maxwellMetricAffine}
(for vanishing cosmological constant cases, see~\citep{Salgado:2014-Topological,Concha:2014-ChernSimons,azcarraga2014minimalsuper,Concha:2014-N1Supergravity,Aviles:2018-Non-Relativistic,Concha:2018-Asymptotic}).
Up to now, the~energy-momentum tensor coming from the Maxwell extension
has not been extensively analyzed yet, but in this context, a~minimal
cosmological model has been introduced in~\citep{durka2011local}
and also it is thought that the gauge fields of the Maxwell symmetry
may provide a geometric background to describe vector inflatons in
cosmological models~\citep{Azcarraga2013maxwellApplication} (for
different solutions, see~\citep{hoseinzadeh20142+,Concha:2018-Asymptotic}).
It is well known that such an additional term may be related to dark
energy~\citep{frieman2008dark,padmanabhan2009dark}. For~the non-gravitational
case, Maxwell symmetry is also used to describe planar dynamics of
the Landau problem~\citep{fedoruk2012new}, higher spin fields~\citep{fedoruk2013maxwell,fedoruk2013new},
and applied to the string theory as an internal symmetry of the matter
gauge fields~\citep{hoseinzadeh20142+}. Also, recent papers~\citep{Obukhov:2021AlgebraOfSymmetry,Obukhov:2022AlgebrasIntegrals,Obukhov:2022MaxwellsEquations}
{have applied} 
the Maxwell group in a classical form, which is given in~\citep{bacry1970group,schrader1972maxwell}.

Furthermore, it is proposed that the renormalizability and unitarity
problems in quantum gravity can be overcome by taking the affine group
as the dynamical group in the gauge theory of gravity, with the help
of generalized linear connection~\citep{Neeman:1979UnifiedAffine,Neeman:1988GaugeAffine,Neeman:1988HadronSL,Lee:1990RenormalizationAffine,Hehl:1995MetricAffine,Lopez-Pinto:1995NonLinearAffineGauge}.
In this paper, we examine the Maxwell extension for both general-linear
and special-linear groups in $D$ dimensions and analyze their gauge
theory of gravity, in particular the generalized cosmological constant
term. We have already studied these groups in four space-time dimensions
in~\citep{cebeciouglu2015maxwell,kibarouglu2019maxwellSpecial,cebeciouglu2021maxwellMetricAffine},
thus this work will generalize our previous results to $D$-dimensional
space-time.

The organization of the paper is as follows. In~Section~\ref{sec: 2},
we study the Maxwell extensions of the general linear group $\mathcal{GL}\left(n,\mathbb{R}\right)$.
Then we show that several Maxwell algebras can be obtained when we
choose appropriate subalgebras and reduce it from a $5$-dimensional
case to $4$ dimensions. We also construct the gauge theory of gravity
based on the Maxwell extended $\mathcal{GL}\left(5,\mathbb{R}\right)$.
After applying dimensional reduction from $5$ to $4$~dimensions,
we analyze the gravity action for two different cases. In~Section~\ref{sec: 3},
we present the special-linear group $\mathcal{SL}\left(n+1,\mathbb{R}\right)$
and its Maxwell extension. In~this framework, we give three examples
that this extension leads to derive different generalizations of the
Maxwell algebra in $4$ dimensions. Similar to the previous section,
we present the gauge theory of gravity based on this extended case.
Finally, Section~\ref{sec: 4} concludes the paper with some~discussions.

\section{Maxwell Extensions of \boldmath{$\mathcal{GL}\left(n,\mathbb{R}\right)$}
Group\label{sec: 2}}

The general linear group $\mathcal{GL}\left(n,\mathbb{R}\right)$
which corresponds to the set of all linear transformations satisfies
the following Lie algebra, 
\begin{eqnarray}
\left[\mathcal{L}_{\,\,B}^{A},\mathcal{L}_{\,\,D}^{C}\right] & = & i\left(\delta_{\,\,B}^{C}\mathcal{L}_{\,\,D}^{A}-\delta{}_{\,\,D}^{A}\mathcal{L}{}_{\,\,B}^{C}\right),
\end{eqnarray}
where $\mathcal{L}_{\,\,B}^{A}$ are the generators of the group.
This algebra can be extended to Maxwell-general linear algebra by
adding an anti-symmetric tensor generator, $\mathcal{\mathcal{Z}}_{AB}$,
associated with the Maxwell symmetry. The~generators $\mathcal{L}_{\,\,B}^{A}$
and $\mathcal{\mathcal{Z}}_{AB}$ obey the commutators 
\begin{eqnarray}
\left[\mathcal{L}_{\,\,B}^{A},\mathcal{L}_{\,\,D}^{C}\right] & = & i\left(\delta_{\,\,B}^{C}\mathcal{L}_{\,\,D}^{A}-\delta{}_{\,\,D}^{A}\mathcal{L}{}_{\,\,B}^{C}\right),\nonumber \\
\left[\mathcal{L}_{\,\,B}^{A},\mathcal{\mathcal{Z}}_{CD}\right] & = & i\left(\delta_{\,\,D}^{A}\mathcal{Z}_{BC}-\delta_{\,\,C}^{A}\mathcal{Z}_{BD}\right),\nonumber \\
\left[\mathcal{Z}_{AB},\mathcal{Z}_{CD}\right] & = & 0,\label{eq: mgl5}
\end{eqnarray}
where the capital Latin indices run $A,B,C,...=0,1,...,n-1$ and $n$
is the dimension of the group. The~algebra with the commutation relations
given by Equation~(\ref{eq: mgl5}), is denoted as $\mathcal{MGL}\left(n,\mathbb{R}\right)$,
the Maxwell-general linear~algebra.

\subsection{Decomposition of $\mathcal{MGL}\left(5,\mathbb{R}\right)$}

In this section, we start with the 35-dimensional $\mathcal{MGL}\left(5,\mathbb{R}\right)$
algebra in the 5-dimensional space-time and carry out the dimensional
reduction to the 4 dimensions. For~this purpose, if~we define the
following generators 
\begin{equation}
L_{\,\,b}^{a}=\mathcal{L}_{\,\,b}^{a},\,\,\,\,\,P_{a}=\left(\mathcal{L}_{\,\,a}^{4}-\frac{\lambda}{2}\mathcal{Z}_{4a}\right),\,\,\,\,\,\,Z_{ab}=\mathcal{Z}_{ab},
\end{equation}
and this definition yields the 26-dimensional subalgebra of $\mathcal{MGL}\left(5,\mathbb{R}\right)$
as, 
\begin{eqnarray}
\left[L_{\,\,b}^{a},L_{\,\,d}^{c}\right] & = & i\left(\delta_{\,\,b}^{c}L_{\,\,d}^{a}-\delta{}_{\,\,d}^{a}L{}_{\,\,b}^{c}\right),\nonumber \\
\left[L_{\,\,b}^{a},P_{c}\right] & = & -i\delta_{\,\,c}^{a}P_{b},\nonumber \\
\left[P_{a},P_{b}\right] & = & i\lambda Z_{ab},\nonumber \\
\left[L_{\,\,b}^{a},Z_{cd}\right] & = & i\left(\delta_{\,\,d}^{a}Z_{bc}-\delta_{\,\,c}^{a}Z_{bd}\right),\label{eq: mga(4,R)}
\end{eqnarray}
where the generators $L_{\,\,b}^{a}$, $P_{a}$, $Z_{ab}$ correspond
to the general linear transformation, translation, and Maxwell symmetry
transformation, respectively. Here, the~constant $\lambda$ has the
unit of $L^{-2}$ which will be related to the cosmological constant
where $L$ is considered as the unit of length. We note that the small
Latin indices are $a,b,c,...=0,1,2,3$ and the remaining commutators
are zero. This algebra is the Maxwell extension of general affine
$\mathcal{GA}\left(4,\mathbb{R}\right)$ algebra which is the semi-direct
product of the general-linear group $\mathcal{GL}\left(4,\mathbb{R}\right)$
with the group of translation $T_{4}$ (for more details, see~\citep{Hehl:1977HadronDilatation,Lord:1978MetricAffine,Neeman:1979UnifiedAffine}).
This 26-dimensional extended group is denoted by $\mathcal{MGA}\left(4,\mathbb{R}\right)$.
The method of nonlinear realization~\citep{Coleman:1969-1,Callan:1969-2,Salam:1969nonlinear1,Salam:1969nonlinear2},
allows us to obtain a differential realization of the generators as
\citep{cebeciouglu2015maxwell} 
\begin{eqnarray}
P_{a} & = & i\left(\partial_{a}-\frac{\lambda}{2}x^{b}\partial_{ab}\right),\nonumber \\
Z_{ab} & = & i\partial_{ab},\nonumber \\
L_{\,\,b}^{a} & = & i\left(x^{a}\partial_{b}+2\theta^{ac}\partial_{bc}\right),
\end{eqnarray}
where $\partial_{a}=\frac{\partial}{\partial x^{a}}$, $\partial_{ab}=\frac{\partial}{\partial\theta^{ab}}$,
and $\partial_{ab}\theta^{cd}=\frac{1}{2}\left(\delta_{a}^{c}\delta_{b}^{d}-\delta_{b}^{c}\delta_{a}^{d}\right)$.
One can check that these differential operators fulfill the Maxwell--affine
algebra and verify the self-consistency of Jacobi identities.

Taking the tangent space (Minkowski metric), $\eta_{ab}=diag\left(+,-,-,-\right)$
into consideration, we can define the following generators, 
\begin{equation}
M_{ab}=\eta_{[ac}L_{\,\,b]}^{c},\,\,\,\,\,T_{ab}=\eta_{(ac}L_{\,\,b)}^{c},\,\,\,\,\,P_{a}=\left(\mathcal{L}_{\,\,a}^{4}-\frac{\lambda}{2}\mathcal{Z}_{4a}\right),\,\,\,\,\,\,Z_{ab}=\mathcal{Z}_{ab},
\end{equation}
where the antisymmetrization and symmetrization of the objects are
defined by $A_{[a}B_{b]}=A_{a}B_{b}-A_{b}B_{a}$ and $A_{(a}B_{b)}=A_{a}B_{b}+A_{b}B_{a}$,
respectively. Thus, the~Lie algebra of these generators can be given
as, 
\begin{eqnarray}
\left[M_{ab},M_{cd}\right] & = & i\left(\eta_{ad}M_{bc}+\eta_{bc}M_{ad}-\eta_{ac}M_{bd}-\eta_{bd}M_{ac}\right),\nonumber \\
\left[M_{ab},T_{cd}\right] & = & i\left(-\eta_{ad}T_{bc}+\eta_{bc}T_{ad}-\eta_{ac}T_{bd}+\eta_{bd}T_{ac}\right),\nonumber \\
\left[T_{ab},T_{cd}\right] & = & i\left(\eta_{ad}M_{bc}+\eta_{bc}M_{ad}+\eta_{ac}M_{bd}-\eta_{bd}M_{ac}\right),\nonumber \\
\left[M_{ab},P_{c}\right] & = & i\left(\eta_{bc}P_{a}-\eta_{ac}P_{b}\right),\nonumber \\
\left[T_{ab},P_{c}\right] & = & -i\left(\eta_{bc}P_{a}+\eta_{ac}P_{b}\right),\nonumber \\
\left[P_{a},P_{b}\right] & = & i\lambda Z_{ab},\nonumber \\
\left[M_{ab},Z_{cd}\right] & = & i\left(\eta_{ad}Z_{bc}+\eta_{bc}Z_{ad}-\eta_{ac}Z_{bd}-\eta_{bd}Z_{ac}\right),\nonumber \\
\left[T_{ab},Z_{cd}\right] & = & i\left(\eta_{ad}Z_{bc}-\eta_{bc}Z_{ad}-\eta_{ac}Z_{bd}+\eta_{bd}Z_{ac}\right),\label{eq: mga4_long}
\end{eqnarray}
where $M_{ab}$ is the anti-symmetric Lorentz generator, $T_{ab}$
is the symmetric deformation generator, $P_{a}$ is the translation
generator, and $Z_{ab}$ is the Maxwell symmetry generator. The~algebra
spanned by \{$M_{ab}$,$T_{ab}$,$P_{a}$, $Z_{ab}$\} is the Maxwell--affine
algebra introduced in Equation~(\ref{eq: mga(4,R)}). This is also
a minimal Maxwell extension of $\mathcal{GA}\left(4,\mathbb{R}\right)$
group. The differential realization of the generators are 
\begin{eqnarray}
M_{ab} & = & i\left(x_{[a}\partial_{b]}+2\theta_{[a}^{\,\,c}\partial_{b]c}\right),\nonumber \\
T_{ab} & = & i\left(x_{(a}\partial_{b)}+2\theta_{(a}^{\,\,c}\partial_{b)c}\right),\nonumber \\
P_{a} & = & i\left(\partial_{a}-\frac{\lambda}{2}x^{b}\partial_{ab}\right),\nonumber \\
Z_{ab} & = & i\partial_{ab}.
\end{eqnarray}

Moreover, if~we consider the following definitions with the Minkowski
metric $\eta_{ab}$, 
\begin{equation}
M_{ab}=\eta_{[ac}\mathcal{L}_{\,\,b]}^{c},\,\,\,\,\,P_{a}=\left(\mathcal{L}_{\,\,a}^{4}-\frac{\lambda}{2}\mathcal{Z}_{4a}\right),\,\,\,\,\,D=\mathcal{L}_{\,\,4}^{4}\,\,\,\,\,\,Z_{ab}=\mathcal{Z}_{ab},
\end{equation}
then we get the 17-dimensional subalgebra with following non-zero
commutation relations, 
\begin{eqnarray}
\left[M_{ab},M_{cd}\right] & = & i\left(\eta_{ad}M_{bc}+\eta_{bc}M_{ad}-\eta_{ac}M_{bd}-\eta_{bd}M_{ac}\right),\nonumber \\
\left[M_{ab},P_{c}\right] & = & i\left(\eta_{bc}P_{a}-\eta_{ac}P_{b}\right),\nonumber \\
\left[P_{a},P_{b}\right] & = & i\lambda Z_{ab},\nonumber \\
\left[M_{ab},Z_{cd}\right] & = & i\left(\eta_{ad}Z_{bc}+\eta_{bc}Z_{ad}-\eta_{ac}Z_{bd}-\eta_{bd}Z_{ac}\right),\nonumber \\
\left[P_{a},D\right] & = & iP_{a},\nonumber \\
\left[Z_{ab},D\right] & = & 2iZ_{ab}.\label{eq: mw}
\end{eqnarray}

This is the Maxwell--Weyl algebra introduced in~\citep{Bonanos:2010DeformationsMaxwellSuper}
and studied in the context of the gauge theory of gravity in~\citep{cebeciouglu2014gauge}.
The differential realization of the generators \{$M_{ab}$,$P_{a},D$,$Z_{ab}$\}
is given by 
\begin{eqnarray}
M_{ab} & = & i\left(x_{[a}\partial_{b]}+2\theta_{[a}^{\,\,c}\partial_{b]c}\right),\nonumber \\
P_{a} & = & i\left(\partial_{a}-\frac{\lambda}{2}x^{b}\partial_{ab}\right),\nonumber \\
D & = & i\left(x^{a}\partial_{a}+2\theta^{ab}\partial_{ab}\right),\nonumber \\
Z_{ab} & = & i\partial_{ab},
\end{eqnarray}
and they correspond to the generalized Lorentz transformations, space-time
translations, dilatation, and the Maxwell symmetry transformations,
respectively. In~the absence of the dilatation symmetry, $\mathcal{L}_{\,\,4}^{4}=0$,
the algebra in Equation~(\ref{eq: mw}) reduces to the well-known
Maxwell algebra~\citep{soroka2005tensor,azcarraga2011generalized}.
Herein we can conclude that $\mathcal{MGL}\left(n,\mathbb{R}\right)$
algebra is a comprehensive symmetry algebra which has the great potential
for obtaining different kinds of Maxwell extended algebras.

\subsection{Gauging the $\mathcal{MGL}\left(5,\mathbb{R}\right)$ Algebra}

In this part, we consider the gauge theory of the $\mathcal{MGL}\left(5,\mathbb{R}\right)$
algebra. To~gauge this algebra, we follow the methods presented in
\citep{azcarraga2011generalized,cebeciouglu2015maxwell}. At~first,
we need to introduce gauge potentials, i.e.,~a vector-valued 1-form
$\mathcal{A}\left(x\right)=\mathcal{A}^{\mathcal{A}}\left(x\right)X_{\mathcal{A}}$
as 
\begin{equation}
\mathcal{A}\left(x\right)=\tilde{\omega}_{\,\,A}^{B}\mathcal{L}_{\,\,B}^{A}+B^{AB}\mathcal{Z}_{AB},
\end{equation}
where $\mathcal{A}^{\mathcal{A}}\left(x\right)=\left\{ \tilde{\omega}_{\,\,A}^{B},B^{AB}\right\} $
are the gauge fields corresponding to the generators $X_{\mathcal{A}}=\left\{ \mathcal{L}_{\,\,B}^{A},\mathcal{Z}_{AB}\right\} $.
The variation of the gauge field $\mathcal{A}\left(x\right)$ under
infinitesimal gauge transformation in tangent space can be calculated
by using the following formula 
\begin{equation}
\delta\mathcal{A}=-d\zeta-i\left[\mathcal{A},\zeta\right],\label{eq: gauge_fields_var}
\end{equation}
with the gauge generator 
\begin{equation}
\zeta(x)=\tilde{\lambda}_{\,\,A}^{B}\mathcal{L}_{\,\,B}^{A}+\varphi^{AB}\mathcal{Z}_{AB},
\end{equation}
where, $\tilde{\lambda}_{\,\,A}^{B}(x)$ and $\varphi^{AB}(x)$, are
the parameters of the corresponding generators. The~transformation
properties of the gauge fields under infinitesimal action of the $\mathcal{MGL}\left(5,\mathbb{R}\right)$
are 
\begin{eqnarray}
\delta\tilde{\omega}_{\,\,B}^{A} & = & -d\tilde{\lambda}_{\,\,B}^{A}+\tilde{\lambda}_{\,\,B}^{C}\omega_{\,\,C}^{A}-\tilde{\lambda}_{\,\,C}^{A}\omega_{\,\,B}^{C},\nonumber \\
\delta B^{AB} & = & -d\varphi^{AB}+\tilde{\lambda}_{\,\,\,C}^{[A}B^{CB]}-\tilde{\omega}_{\,\,\,C}^{[A}\varphi^{CB]}.
\end{eqnarray}

The curvature 2-form $\mathcal{F}\left(x\right)$ is given by the
structure equation 
\begin{equation}
\mathcal{F}=d\mathcal{A}+\frac{i}{2}\left[\mathcal{A},\mathcal{A}\right],\label{eq: curvatures}
\end{equation}
written in terms of components 
\begin{equation}
\mathcal{F}\left(x\right)=\tilde{\mathcal{R}}_{\,\,A}^{B}\mathcal{L}_{\,\,B}^{A}+\mathcal{F}^{AB}\mathscr{\mathcal{Z}}_{AB},
\end{equation}
one can calculate the curvature 2-forms of the associated gauge fields
as 
\begin{eqnarray}
\tilde{\mathcal{R}}_{\,\,B}^{A} & = & d\tilde{\omega}_{\,\,B}^{A}+\tilde{\omega}_{\,\,C}^{A}\wedge\tilde{\omega}_{\,\,B}^{C}=\mathcal{D}\tilde{\omega}_{\,\,B}^{A},\nonumber \\
\mathcal{F}^{AB} & = & dB^{AB}+\tilde{\omega}_{\,\,\,C}^{[A}\wedge B^{CB]}=\mathcal{D}B^{AB},
\end{eqnarray}
where $\mathcal{D}=d+\tilde{\omega}$ is the exterior covariant derivative
with respect to $\mathcal{MGL}\left(5,\mathbb{R}\right)$ connection
$\tilde{\omega}_{\,\,B}^{A}$. These is the $\mathcal{GL}\left(5,\mathbb{R}\right)$
curvature 2-form and a new curvature 2-form for the tensor generator
$\mathcal{Z}_{AB}$. Under an infinitesimal gauge transformation with
parameters $\zeta(x)$, the change in curvature is given by 
\begin{equation}
\delta\mathcal{F}=i\left[\zeta,\mathcal{F}\right],\label{eq: curvatures_var}
\end{equation}
and hence one gets 
\begin{eqnarray}
\delta\tilde{\mathcal{R}}_{\,\,B}^{A} & = & \tilde{\lambda}_{\,\,C}^{A}\tilde{\mathcal{R}}_{\,\,B}^{C}-\tilde{\lambda}_{\,\,B}^{C}\tilde{\mathcal{R}}_{\,\,C}^{A},\nonumber \\
\delta\mathcal{F}^{AB} & = & \tilde{\lambda}_{\,\,\,C}^{[A}\mathcal{F}^{CB]}-\tilde{\mathcal{R}}_{\,\,\,C}^{[A}\varphi^{CB]},\label{eq: var_curv}
\end{eqnarray}
the gauge variations of the curvatures. By~taking the exterior covariant
derivatives of the curvatures, the~Bianchi identities become 
\begin{eqnarray}
\mathcal{D}\tilde{\mathcal{R}}_{\,\,B}^{A} & = & 0,\nonumber \\
\mathcal{D}\mathcal{F}^{AB} & = & \tilde{\mathcal{R}}{}_{\,\,\,C}^{[A}B^{CB]}.
\end{eqnarray}

Having found the transformations of the gauge fields and the curvatures,
we are ready to look for an invariant gravitational Lagrangian under
these~transformations.

\subsection{Gravitational~Action}

In this section, we follow the approach of Stelle and West (SW)~\citep{Stelle:1979va,Stelle:1979aj}.
Our starting point is the local $\mathcal{SO}\left(2,3\right)$ symmetry
with the metric signature as $\left(+,-,-,-,+\right)$ on a 4-dimensional
Minkowski space-time. It is well known that the symmetry-breaking
mechanism of the Stelle--West model provides a physically realistic
mechanism for obtaining gravity as a gauge theory with a spontaneously
broken local symmetry. Stelle and West considered an action where
a symmetry-breaking mechanism is induced by introducing a non-dynamical
vector field $V^{A}$ in order to promote local $\mathcal{SO}\left(2,3\right)$
transformations to gauge symmetries, which is constrained by the condition
$V_{A}V^{A}=c^{2}$.

The SW action can be given by 
\begin{equation}
S_{SW}=\sigma\int V^{E}\epsilon_{ABCDE}\mathcal{R}^{AB}\wedge\mathcal{R}^{CD}+\alpha\left(c^{2}-V_{A}V^{A}\right),\label{eq: action_SW}
\end{equation}
where $\sigma$ is a constant and $\alpha$ is an arbitrary 4-form
serving as a Lagrange multiplier. The~totally anti-symmetric symbol
$\epsilon_{ABCDE}$ is an invariant tensor of the algebra $so\left(2,3\right)$.
We can also note that $\mathcal{A}^{AB}\left(x\right)$ is a de-Sitter
connection 1-form, and~$\mathcal{R}^{AB}$$\left(x\right)$ is its
curvature 2-form. Choosing 
\begin{equation}
V^{A}=\left(c,0,0,0,0\right),
\end{equation}
\begin{equation}
e^{a}\left(x\right)=-lDV^{a}=-lc\mathcal{A}_{4}^{a},\,\,\,\,\,\,\,DV^{4}=0,\label{eq: SW_vierbein}
\end{equation}
where $e^{a}\left(x\right)$ corresponds to the vierbein field, $D$
is the Lorentz covariant derivative and $l$ is related to the cosmological
constant according to $l=\sqrt{3/\Lambda}$, $\mathcal{SO}\left(2,3\right)$
symmetry is broken spontaneously to $\mathcal{SO}\left(1,3\right)$,
and we obtain the action 
\begin{equation}
S_{SW}=\frac{1}{2\kappa}\int\epsilon_{abcd}\left(R^{ab}\wedge e^{c}\wedge e^{d}-\frac{\Lambda}{6}e^{a}\wedge e^{b}\wedge e^{c}\wedge e^{d}\right),
\end{equation}
where $R^{ab}=d\omega^{ab}+\omega_{\,\,c}^{a}\wedge\omega^{cb}$ is
the Ricci curvature 2-form, and~we identify $\mathcal{A}^{ab}\left(x\right)=\omega^{ab}\left(x\right)$
and also set $\sigma c=-\frac{3}{4\kappa\Lambda}$ together with $\kappa$
is being Einstein's gravitational~constant.

In analogy with the Stelle--West action in Equation~(\ref{eq: action_SW}),
we will try to construct a gravitational action generalized to the
case of the $\mathcal{MGL}\left(5,\mathbb{R}\right)$ symmetry group.
For this purpose, we combine the curvature 2-forms $\tilde{\mathcal{R}}_{\,\,B}^{A}$
and $\mathcal{F}^{AB}$ into an anti-symmetric shifted curvature 2-form
as follows 
\begin{equation}
\mathcal{J}^{AB}=\tilde{\mathcal{R}}^{AB}-\mu\mathcal{F}^{AB},\label{eq: J}
\end{equation}
where $\mu$ is a dimensionless constant which will be employed in
the definition of the cosmological constant $\Lambda=\mu\lambda$
and we define a new object as $\tilde{\mathcal{R}}^{AB}=\tilde{\mathcal{R}}_{\,\,C}^{[A}g^{CB]}$.
Here we introduced an additional symmetric tensor field $g^{AB}$,
which is called the premetric tensor field. It does not represent
the metric tensor of space-time but it helps to construct an invariant
Lagrangian (for more details, see~\citep{cebeciouglu2015maxwell,Hehl:1991_CS}).
The components of the premetric field are $\mathcal{GL}\left(5,\mathbb{R}\right)$
tensor valued $0$-forms and the infinitesimal transformation of the
premetric tensor field under the local $\mathcal{GL}\left(5,\mathbb{R}\right)$
symmetry group is 
\begin{equation}
\delta g^{AB}=\tilde{\lambda}_{\,\,\,C}^{(A}g^{CB)},\,\,\,\,\,\delta g_{AB}=-\tilde{\lambda}_{\,\,(A}^{C}g_{CB)},\label{eq: premetric_trans}
\end{equation}
and its covariant derivative is given as 
\begin{equation}
\mathcal{Q}^{AB}=\mathcal{D}g^{AB}=dg^{AB}+\tilde{\omega}_{\,\,\,C}^{(A}\wedge B^{CB)}.
\end{equation}

We note that if one identifies $g^{AB}$ as the space-time metric
tensor, then $\mathcal{Q}^{AB}$ becomes the nonmetricity. On~the
other hand, this kind of additional metric-like fields may be discussed
in the context of the metric-affine gravity~\citep{Hehl:1995MetricAffine,Hohmann:2019Metric-affine,cebeciouglu2021maxwellMetricAffine}.

By using Equation~(\ref{eq: premetric_trans}), we can find the gauge
transformation of the new object as $\delta\tilde{\mathcal{R}}^{AB}=\tilde{\lambda}_{\,\,\,C}^{[A}\tilde{\mathcal{R}}^{CB]}$.
Moreover, since the tensorial translation being traded for diffeomorphism
invariance is not {symmetric} 
of the action~\citep{durka2011gauged}, omitting the tensorial-space
translations, the~transformation rules for the curvature $\mathcal{F}^{AB}$
in Equation~(\ref{eq: var_curv}) can be rewritten as $\delta\mathcal{F}^{AB}=\tilde{\lambda}_{\,\,\,C}^{[A}\mathcal{F}^{CB]}$.
Then using this background, one can show that the shifted curvature
in Equation~(\ref{eq: J}) has the following gauge transformation,
\begin{equation}
\delta\mathcal{J}^{AB}=\tilde{\lambda}_{\,\,\,C}^{[A}\mathcal{J}^{CB]}.\label{eq: var_J}
\end{equation}

Now, we can start with the following generalized Stelle--West action
\begin{equation}
S_{MGL}=\sigma\int V^{E}\eta_{ABCDE}\mathcal{J}^{AB}\wedge\mathcal{J}^{CD}+\alpha\left(c^{2}-V_{A}V^{A}\right),\label{eq: action_1_mgl}
\end{equation}
where $V^{A}$ is a 0-form non-dynamical $\mathcal{GL}\left(5,\mathbb{R}\right)$
five-vector field with dimensions of the length and it satisfies the
transformation $\delta V^{A}=\tilde{\lambda}_{\,\,\,C}^{A}V^{C}$.
We also use another definition $V_{A}=V^{B}g_{AB}$ which obeys the
transformation rule as $\delta V_{A}=-\tilde{\lambda}_{\,\,\,A}^{C}V_{C}$.
Note that the totally anti-symmetric symbol $\eta_{ABCDE}$ is an
invariant tensor under $gl\left(5,\mathbb{R}\right)$ algebra with
the following transformation rule, 
\begin{equation}
\delta\eta_{ABCDE}=-\tilde{\lambda}_{\,\,\,A}^{F}\eta_{FBCDE}-\tilde{\lambda}_{\,\,\,B}^{F}\eta_{AFCDE}-\tilde{\lambda}_{\,\,\,C}^{F}\eta_{ABFDE}-\tilde{\lambda}_{\,\,\,D}^{F}\eta_{ABCFE}-\tilde{\lambda}_{\,\,\,E}^{F}\eta_{ABCDF}\label{eq: eta_trans}
\end{equation}

With the help of Equation~(\ref{eq: var_J}), one can easily check
that the action Equation~(\ref{eq: action_1_mgl}) is gauge invariant.
We note that $\mathcal{J}^{AB}$ is an asymmetric under the interchange
of indices due to the definition of $\tilde{\mathcal{R}}^{AB}.$ Thus,
only the anti-symmetric part of $\tilde{\mathcal{R}}^{AB}$ contributes
to the gravitational dynamics in the action (\ref{eq: action_1_mgl}).

Then by the variation of the action with respect to $\tilde{\omega}_{\,\,F}^{A}\left(x\right)$,
$B^{AB}\left(x\right)$ and $g^{FG}\left(x\right)$, we obtain the
field equations 
\begin{eqnarray}
\mathcal{D}\left(\eta_{ABCDE}V^{E}g^{FB}\mathcal{J}^{CD}\right)-\mu\eta_{ABCDE}V^{E}B^{FB}\wedge\mathcal{J}^{CD} & = & 0,\nonumber \\
\mathcal{D}\left(V^{E}\eta_{ABCDE}\mathcal{J}^{CD}\right) & = & 0,\label{eq: eom_mgl5}\\
V^{E}\eta_{ACDE(F}\tilde{\mathcal{R}}{}_{\,\,G)}^{A}\wedge\mathcal{J}^{CD}+\frac{1}{2}V^{E}g_{FG}\eta_{ABCDE}\mathcal{J}^{AB}\wedge\mathcal{J}^{CD} & = & 0,\nonumber 
\end{eqnarray}
and they are invariant under local $\mathcal{MGL}\left(5,\mathbb{R}\right)$
transformations.

\subsubsection*{Four-Dimensional~Case}

If we take $V^{A}=\left(c,0,0,0,0\right)$ and fix the constant as
$\sigma c=-\frac{3}{4\kappa\Lambda}$, the~gravitational action Equation~(\ref{eq: action_1_mgl})
spontaneously breaks down to, 
\begin{eqnarray}
S_{MGL} & = & -\frac{3}{4\kappa\Lambda}\int\eta_{abcd}\mathcal{J}^{ab}\wedge\mathcal{J}^{cd}\nonumber \\
 & = & -\frac{3}{4\kappa\Lambda}\int\eta_{abcd}\left(\tilde{\mathcal{R}}_{\,\,e}^{[a}g^{eb]}\wedge\tilde{\mathcal{R}}_{\,\,f}^{[c}g^{fd]}-2\mu\tilde{\mathcal{R}}_{\,\,e}^{[a}g^{eb]}\wedge\mathcal{F}^{cd}+\mu^{2}\mathcal{F}^{ab}\wedge\mathcal{F}^{cd}\right),\label{eq: action_mga(4R)}
\end{eqnarray}
where $\eta_{abcd}=e\epsilon_{abcd}$ obey the same transformation
rule given in Equation~(\ref{eq: eta_trans}) and $e$ is the determinant
of the vierbein field and $\epsilon_{abcd}$ is the Levi--Civita
symbol. This action has a structural similarity to the Maxwell--Affine
gravity action which was given in~\citep{cebeciouglu2015maxwell}
but includes more general curvature 2-forms. So we can say that if
we construct the Stelle--West-like action by using curvature 2-forms
which come from $\mathcal{MGL}\left(5,\mathbb{R}\right)$ we can obtain
a generalized framework for Maxwell gravity. Furthermore, if~we assume
the premetric field $g^{ab}\left(x\right)$ as the tangent space metric
tensor, then the action (\ref{eq: action_1_mgl}) can be written as
follows 
\begin{eqnarray}
S_{MGL} & = & -\frac{3}{4\kappa\Lambda}\int\eta_{abcd}\mathcal{J}^{ab}\wedge\mathcal{J}^{cd}\nonumber \\
 & = & -\frac{3}{4\kappa\Lambda}\int\eta_{abcd}\left(\tilde{\mathcal{R}}^{ab}\wedge\tilde{\mathcal{R}}^{cd}-2\mu\tilde{\mathcal{R}}^{ab}\wedge\mathcal{F}^{cd}+\mu^{2}\mathcal{F}^{ab}\wedge\mathcal{F}^{cd}\right).\label{eq: action_MW}
\end{eqnarray}
where $\tilde{\mathcal{R}}^{ab}$ and $\mathcal{F}^{ab}$ are anti-symmetric
curvature 2-forms. This action generalizes the minimal Maxwell gravity
which was discussed in~\citep{azcarraga2011generalized}. Let us
expand this action to a more explicit form. Similar to the definition
in Equation~(\ref{eq: SW_vierbein}), we define the following vector
fields, 
\begin{equation}
e^{a}\left(x\right)=-\sqrt{\frac{3}{\Lambda}}\omega^{a5},\,\,\,\,b^{a}\left(x\right)=-\sqrt{\frac{3}{\Lambda}}B^{a5}
\end{equation}
where $e^{a}\left(x\right)$ can be considered as the vierbein vector
field and $b^{a}\left(x\right)$ is an additional vector field as
an effect of the Maxwell symmetry. Then the shifted curvature becomes
\begin{equation}
\mathcal{J}^{ab}=R^{ab}\left(\omega\right)-\mu DB^{ab}-\frac{\Lambda}{3}\left(e^{a}\wedge e^{b}-\mu e^{[a}\wedge b^{b]}\right)
\end{equation}
where $D$ is the Lorentz covariant derivative and $R^{ab}\left(\omega\right)$
is the Riemann curvature 2-form. From~this background, neglecting
the total derivatives, the~action (\ref{eq: action_MW}) reduces
\begin{eqnarray}
S_{MGL} & = & \frac{1}{2\kappa}\int\eta_{abcd}\left(R^{ab}\wedge e^{c}\wedge e^{d}-\frac{\Lambda}{6}e^{a}\wedge e^{b}\wedge e^{c}\wedge e^{d}\right)\nonumber \\
 &  & -2\mu\eta_{abcd}\left(R^{ab}\wedge e^{c}\wedge b^{d}+\frac{1}{2}DB^{ab}\wedge e^{c}\wedge e^{d}-\mu DB^{ab}\wedge e^{c}\wedge b^{d}\right)\nonumber \\
 &  & +\frac{2\Lambda\mu}{3}\eta_{abcd}\left(e^{a}\wedge e^{b}\wedge e^{c}\wedge b^{d}-\mu e^{a}\wedge b^{b}\wedge e^{c}\wedge b^{d}\right).\label{eq: Action_CC}
\end{eqnarray}

The first line includes the Einstein--Hilbert-like term together
with the cosmological constant term, the~second line contains mixed
terms, and the last line corresponds to the generalized cosmological
terms with the Maxwell symmetry~contributions.

Thus we can say that the gauge theory of gravity based on $\mathcal{MGL}\left(5,\mathbb{R}\right)$
extends the geometrical framework of Einstein's gravitational theory
to be included the generalized cosmological constant term. So this
theory provides an alternative way to introduce the cosmological term.

\section{Maxwell Extensions of the \boldmath{$\mathcal{SL}\left(n+1,\mathbb{R}\right)$}
Group\label{sec: 3}}

The group of special-linear transformations (also known as the metalinear
group) $\mathcal{SL}\left(n+1,\mathbb{R}\right)$, being the subgroup
of $\mathcal{GL}\left(n,\mathbb{R}\right)$, consists of matrices
of determinant unity and it is generated by $n\left(n+2\right)$ trace-free
generators, 
\begin{equation}
\mathring{\mathcal{L}}_{\,\,B}^{A}=\mathcal{L}_{\,\,B}^{A}-\frac{1}{n}\delta_{B}^{A}\mathcal{L}_{\,\,C}^{C},\label{eq: msl_decompose}
\end{equation}
and satisfies the following Lie algebra~\citep{Mielke:2010_BF,Mielke:2011_Spontaneously_SL5R,Mielke:2012_Einstein_Weyl}
(for more details, see~\citep{Mielke:2017Geometrodynamics_Book}),
\begin{eqnarray}
\left[\mathring{\mathcal{L}}_{\,\,B}^{A},\mathring{\mathcal{L}}_{\,\,D}^{C}\right] & = & i\left(\delta_{\,\,B}^{C}\mathring{\mathcal{L}}_{\,\,D}^{A}-\delta{}_{\,\,D}^{A}\mathring{\mathcal{L}}{}_{\,\,B}^{C}\right),
\end{eqnarray}
where $\mathring{\mathcal{L}}_{\,\,B}^{A}$ are the $\mathcal{SL}\left(n+1,\mathbb{R}\right)$
generators and the index structure is essentially the same as that
of the previous~section.

We extend this algebra by an anti-symmetric tensor generator $\mathcal{\mathcal{Z}}_{AB}$
associated with the Maxwell symmetry as, 
\begin{eqnarray}
\left[\mathring{\mathcal{L}}_{\,\,B}^{A},\mathring{\mathcal{L}}_{\,\,D}^{C}\right] & = & i\left(\delta_{\,\,B}^{C}\mathring{\mathcal{L}}_{\,\,D}^{A}-\delta{}_{\,\,D}^{A}\mathring{\mathcal{L}}{}_{\,\,B}^{C}\right),\nonumber \\
\left[\mathring{\mathcal{L}}_{\,\,B}^{A},\mathcal{\mathcal{Z}}_{CD}\right] & = & i\left(\delta_{\,\,D}^{A}\mathcal{Z}_{BC}-\delta_{\,\,C}^{A}\mathcal{Z}_{BD}+\frac{2}{n}\delta_{\,\,B}^{A}\mathcal{Z}_{CD}\right),\label{eq: msl}\\
\left[\mathcal{Z}_{AB},\mathcal{Z}_{CD}\right] & = & 0.\nonumber 
\end{eqnarray}

This algebra can be named as the Maxwell-special-linear algebra and
denoted with $\mathcal{MSL}\left(n+1,\mathbb{R}\right)$.

\subsection{Decomposition of $\mathcal{MSL}\left(n+1,\mathbb{R}\right)$}

Let us now analyze the algebra (\ref{eq: msl}) in four space-time
dimensions. As~a first example, if~we define the generators as 
\begin{equation}
\mathring{L}_{\,\,b}^{a}=\mathring{\mathcal{L}}_{\,\,b}^{a},\,\,\,\,\,P_{a}=\left(\mathring{\mathcal{L}}_{\,\,a}^{4}-\frac{\lambda}{2}\mathcal{Z}_{4a}\right),\,\,\,\,\,\,Z_{ab}=\mathcal{Z}_{ab},
\end{equation}
then this definition leads to the following 25-dimensional subalgebra
\begin{eqnarray}
\left[\mathring{L}_{\,\,b}^{a},\mathring{L}_{\,\,d}^{c}\right] & = & i\left(\delta_{\,\,b}^{c}\mathring{L}_{\,\,d}^{a}-\delta{}_{\,\,d}^{a}\mathring{L}{}_{\,\,b}^{c}\right),\nonumber \\
\left[\mathring{L}_{\,\,b}^{a},P_{c}\right] & = & -i\left(\delta_{\,\,c}^{a}P_{b}-\frac{1}{4}\delta_{b}^{a}P_{c}\right),\nonumber \\
\left[P_{a},P_{b}\right] & = & i\lambda Z_{ab},\label{eq:msaalg}\\
\left[\mathring{L}_{\,\,b}^{a},Z_{cd}\right] & = & i\left(\delta_{\,\,d}^{a}Z_{bc}-\delta_{\,\,c}^{a}Z_{bd}+\frac{1}{2}\delta_{b}^{a}Z_{cd}\right),\nonumber 
\end{eqnarray}
where the generators $\mathring{L}_{\,\,b}^{a}$, $P_{a}$, $Z_{ab}$
correspond to the generalized special-linear, translation, and the
Maxwell symmetries, respectively. This algebra is the Maxwell extended
special-affine algebra $\mathcal{SA}\left(4,\mathbb{R}\right)$ which
is the semi-direct product of the 15-dimensional $\mathcal{SL}\left(4,\mathbb{R}\right)$
transformation and 4-dimensional translation $T_{4}$. It is also
known as the Maxwell-special-linear algebra and denoted by $\mathcal{MSA}\left(4,\mathbb{R}\right)$
(for more details, see~\citep{kibarouglu2019maxwellSpecial}). The~differential
realization of the generators can also be found as follows 
\begin{eqnarray}
P_{a} & = & i\left(\partial_{a}-\frac{\lambda}{2}x^{b}\partial_{ab}\right),\nonumber \\
Z_{ab} & = & i\partial_{ab},\\
L_{\,\,b}^{a} & = & i\left(x^{a}\partial_{b}+2\theta^{ac}\partial_{bc}\right)-\frac{i}{4}\delta_{\,\,b}^{a}\left(x^{c}\partial_{c}+2\theta^{cd}\partial_{cd}\right),\nonumber 
\end{eqnarray}

As a second example, this time we combine the algebras given in Equations~(\ref{eq: mgl5})
and (\ref{eq: msl}) in four dimensions. Defining the following generators,
\begin{equation}
L_{\,\,b}^{a}=\mathcal{L}_{\,\,b}^{a},\,\,\,\,\,\,\mathring{L}_{\,\,b}^{a}=\mathring{\mathcal{L}}_{\,\,b}^{a},\,\,\,\,\,\,P_{a}=\mathcal{L}_{\,\,a}^{4}-\frac{1}{2}\mathcal{Z}_{4a},\,\,\,\,\,\,P_{*}^{a}=\mathcal{L}_{\,\,4}^{a},\,\,\,\,\,\,Z_{ab}=\mathcal{Z}_{ab},\,\,\,\,\,\,Z_{a}=\mathcal{Z}_{4a},
\end{equation}
we get the Lie algebra as, 
\begin{eqnarray}
\left[L_{\,\,b}^{a},L_{\,\,d}^{c}\right] & = & i\left(\delta_{\,\,b}^{c}L_{\,\,d}^{a}-\delta{}_{\,\,d}^{a}L{}_{\,\,b}^{c}\right),\nonumber \\
\left[L_{\,\,b}^{a},P_{c}\right] & = & -i\delta_{\,\,c}^{a}P_{b},\nonumber \\
\left[L_{\,\,b}^{a},P_{*}^{c}\right] & = & i\delta_{\,\,b}^{c}P_{*}^{a},\nonumber \\
\left[P_{a},P_{b}\right] & = & iZ_{ab},\nonumber \\
\left[P_{*}^{a},P_{*}^{b}\right] & = & 0,\nonumber \\
\left[P_{*}^{a},P_{b}\right] & = & i\left(L_{\,\,b}^{a}-\delta_{\,\,b}^{a}L_{\,\,5}^{5}\right)=i\mathring{L}_{\,\,b}^{a},\label{eq: algebra_metalinear}\\
\left[L_{\,\,b}^{a},Z_{cd}\right] & = & i\left(\delta_{\,\,d}^{a}Z_{bc}-\delta_{\,\,c}^{a}Z_{bd}\right),\nonumber \\
\left[L_{\,\,b}^{a},Z_{c}\right] & = & -i\delta_{\,\,c}^{a}Z_{b},\nonumber \\
\left[P_{a},Z_{cd}\right] & = & 0,\nonumber \\
\left[P_{a},Z_{c}\right] & = & -iZ_{ac},\nonumber \\
\left[P_{*}^{a},Z_{cd}\right] & = & i\left(\delta_{d}^{a}Z_{c}-\delta_{c}^{a}Z_{d}\right),\nonumber \\
\left[P_{*}^{a},Z_{c}\right] & = & 0,\nonumber 
\end{eqnarray}
where we have used the expression $L_{\,\,4}^{4}=\frac{1}{4}L_{\,\,c}^{c}$.
Here, $P_{*}^{a}$ is a vector and $P_{b}$ is a co-vector with respect
to $\mathcal{GL}\left(n,\mathbb{R}\right)$ transformation and they
also generate pseudo-translations. Moreover, there is an additional
vector generator $Z_{a}$ that comes from the Maxwell symmetry extension.
This algebra is a novel Maxwell extension of the meta-linear group
studied in~\citep{Mielke:2011_Spontaneously_SL5R,Mielke:2012_Einstein_Weyl}
in the four-dimensional space-time.

Furthermore, if~the tangent (flat) space carries a metric with the
component $\eta_{ab}$, one can lower the indices and a finer splitting
of $\mathcal{MSA}\left(4,\mathbb{R}\right)$ algebra can be achieved:
\begin{equation}
M_{ab}=\eta_{[ac}\mathring{\mathcal{L}}_{\,\,b]}^{c},\,\,\,\,\,T_{ab}=\eta_{(ac}\mathring{\mathcal{L}}_{\,\,b)}^{c},\,\,\,\,\,P_{a}=\left(\mathring{\mathcal{L}}_{\,\,a}^{4}-\frac{\lambda}{2}\mathcal{Z}_{5a}\right),\,\,\,\,\,\,Z_{ab}=\mathcal{Z}_{ab},
\end{equation}
and the commutation relations given by Equation~(\ref{eq:msaalg})
become 
\begin{eqnarray}
\left[M_{ab},M_{cd}\right] & = & i\left(\eta_{ad}M_{bc}+\eta_{bc}M_{ad}-\eta_{ac}M_{bd}-\eta_{bd}M_{ac}\right),\nonumber \\
\left[M_{ab},T_{cd}\right] & = & i\left(-\eta_{ad}T_{bc}+\eta_{bc}T_{ad}-\eta_{ac}T_{bd}+\eta_{bd}T_{ac}\right),\nonumber \\
\left[T_{ab},T_{cd}\right] & = & i\left(\eta_{ad}M_{bc}+\eta_{bc}M_{ad}+\eta_{ac}M_{bd}-\eta_{bd}M_{ac}\right),\nonumber \\
\left[M_{ab},P_{c}\right] & = & i\left(\eta_{bc}P_{a}-\eta_{ac}P_{b}\right),\nonumber \\
\left[T_{ab},P_{c}\right] & = & -i\left(\eta_{bc}P_{a}+\eta_{ac}P_{b}-\frac{1}{2}\eta_{ab}P_{c}\right),\nonumber \\
\left[P_{a},P_{b}\right] & = & i\lambda Z_{ab},\nonumber \\
\left[M_{ab},Z_{cd}\right] & = & i\left(\eta_{ad}Z_{bc}+\eta_{bc}Z_{ad}-\eta_{ac}Z_{bd}-\eta_{bd}Z_{ac}\right),\nonumber \\
\left[T_{ab},Z_{cd}\right] & = & i\left(\eta_{ad}Z_{bc}-\eta_{bc}Z_{ad}-\eta_{ac}Z_{bd}+\eta_{bd}Z_{ac}+\eta_{ab}Z_{cd}\right),\label{eq: msa4_long}
\end{eqnarray}
where $M_{ab}$ generates the metric-preserving Lorentz subgroup,
$T_{ab}$ generates the (nontrivial) relativistic four-volume-preserving
transformations (shear generator) with $\text{tr}T_{ab}=0$, $P_{a}$
is the translation generator, and $Z_{ab}$ is the Maxwell generator.
The differential realizations of these generators are given by 
\begin{eqnarray}
M_{ab} & = & i\left(x_{[a}\partial_{b]}+2\theta_{[a}^{\,\,c}\partial_{b]c}\right),\nonumber \\
T_{ab} & = & i\left(x_{(a}\partial_{b)}+2\theta_{(a}^{\,\,c}\partial_{b)c}\right)+\frac{i}{2}\eta_{ab}\left(x^{c}\partial_{c}+2\theta^{cd}\partial_{cd}\right),\nonumber \\
P_{a} & = & i\left(\partial_{a}-\frac{\lambda}{2}x^{b}\partial_{ab}\right),\nonumber \\
Z_{ab} & = & i\partial_{ab}.
\end{eqnarray}

Thus, we derived the Maxwell extension of $\mathcal{SA}\left(4,\mathbb{R}\right)$
algebra in the presence of a metric~\citep{Neeman:1979UnifiedAffine}.

\subsection{The Gauge Theory of the $\mathcal{MSL}\left(n+1,\mathbb{R}\right)$
Group}

In the gauging of the $\mathcal{MSL}\left(n+1,\mathbb{R}\right)$
symmetry group, we adopt the same construction procedures given in
the previous section. For~this purpose, we start by writing down
an algebra-valued gauge field 
\begin{equation}
\mathcal{A}\left(x\right)=\mathcal{A}^{A}X_{A}=\mathring{\omega}_{\,\,A}^{B}\wedge\mathring{\mathcal{L}}_{\,\,B}^{A}+B^{AB}\mathcal{Z}_{AB},\label{eq: MSL_A}
\end{equation}
where $\mathcal{A}^{\mathcal{A}}\left(x\right)=\left\{ \mathring{\omega}_{\,\,A}^{B},B^{AB}\right\} $
are the gauge fields corresponding to the generators $X_{\mathcal{A}}=\left\{ \mathring{\mathcal{L}}_{\,\,B}^{A},\mathcal{Z}_{AB}\right\} $,
respectively. Using the Lie algebra valued gauge parameters 
\begin{equation}
\zeta\left(x\right)=\zeta{}^{A}X_{A}=\varphi^{AB}\mathcal{Z}_{AB}+\mathring{\lambda}_{\,\,A}^{B}\wedge\mathring{\mathcal{L}}_{\,\,B}^{A},\label{eq: MSL_A1}
\end{equation}
where $\varphi^{AB}\left(x\right)$ and~$\mathring{\lambda}_{\,\,A}^{B}\left(x\right)$
are the Maxwell and the special-linear transformation parameters,
respectively. By using Equations~(\ref{eq: gauge_fields_var}), (\ref{eq: MSL_A}),
and (\ref{eq: MSL_A1}), the transformation properties of the gauge
fields under the infinitesimal $\mathcal{MSL}\left(n+1,\mathbb{R}\right)$
are 
\begin{eqnarray}
\delta\mathring{\omega}_{\,\,B}^{A} & = & -d\mathring{\lambda}_{\,\,B}^{A}+\mathring{\lambda}_{\,\,B}^{C}\mathring{\omega}_{\,\,C}^{A}-\mathring{\lambda}_{\,\,C}^{A}\mathring{\omega}_{\,\,B}^{C},\nonumber \\
\delta B^{AB} & = & -d\varphi^{AB}+\mathring{\lambda}_{\,\,\,C}^{[A}B^{CB]}-\mathring{\omega}_{\,\,\,C}^{[A}\varphi^{CB]}\nonumber \\
 & = & -d\varphi^{AB}+\tilde{\lambda}_{\,\,\,C}^{[A}B^{CB]}-\tilde{\omega}_{\,\,\,C}^{[A}\varphi^{CB]}-\frac{2}{n}\tilde{\lambda}B^{AB}+\frac{2}{n}\tilde{\omega}\varphi^{AB},
\end{eqnarray}
where $\tilde{\lambda}$ and $\tilde{\omega}$ are the trace parts
of the $\mathcal{GL}\left(n,\mathbb{R}\right)$ valued parameter and
gauge field, respectively. To~find the curvature 2-forms, we make
use of Equation~(\ref{eq: curvatures}) and recall the following~definition
\begin{equation}
\mathcal{F}\left(x\right)=\mathcal{F}^{A}X_{A}=\mathring{\mathcal{R}}_{\,\,A}^{B}\wedge\mathring{\mathcal{L}}_{\,\,B}^{A}+\mathcal{F}^{AB}\mathcal{Z}_{AB}.
\end{equation}

Then it yields the curvature 2-forms of the $\mathcal{MSL}\left(n+1,\mathbb{R}\right)$
algebra as 
\begin{eqnarray}
\mathring{\mathcal{R}}_{\,\,B}^{A} & = & d\mathring{\omega}_{\,\,B}^{A}+\mathring{\omega}_{\,\,C}^{A}\wedge\mathring{\omega}_{\,\,B}^{C}\nonumber \\
 & = & \mathcal{D}\mathring{\omega}_{\,\,B}^{A},\nonumber \\
\mathcal{F}^{AB} & = & dB^{AB}+\mathring{\omega}_{\,\,\,C}^{[A}\wedge B^{CB]}\nonumber \\
 & = & dB^{AB}+\tilde{\omega}_{\,\,\,C}^{[A}\wedge B^{CB]}-\frac{2}{n}\tilde{\omega}B^{AB}\nonumber \\
 & = & \mathcal{D}B^{AB},
\end{eqnarray}
where the exterior covariant derivative $\mathcal{D}$ is defined
with respect to $\mathcal{SL}\left(n+1,\mathbb{R}\right)$ connection.
The infinitesimal variation of the curvature 2-forms under the gauge
transformations can be obtained by using Equation~(\ref{eq: curvatures_var}),
\begin{eqnarray}
\delta\mathring{\mathcal{R}}{}_{\,\,B}^{A} & = & \mathring{\lambda}_{\,\,C}^{A}\mathring{\mathcal{R}}_{\,\,B}^{C}-\mathring{\lambda}_{\,\,B}^{C}\mathring{\mathcal{R}}_{\,\,C}^{A}\nonumber \\
\delta\mathcal{F}{}^{AB} & = & \mathring{\lambda}_{\,\,\,C}^{[A}\mathcal{F}^{CB]}-\mathring{\mathcal{R}}_{\,\,\,C}^{[A}\varphi^{CB]},\nonumber \\
 & = & \tilde{\lambda}_{\,\,\,C}^{[A}\mathcal{F}^{CB]}-\mathring{\mathcal{R}}_{\,\,\,C}^{[A}\varphi^{CB]}+\frac{2}{n}\mathring{\mathcal{R}}\varphi^{AB}-\frac{2}{n}\tilde{\lambda}\mathcal{F}^{AB}.
\end{eqnarray}

As before, we again introduce an additional symmetric tensor field,
$g^{AB}\left(x\right)$ as a premetric tensor field. The~components
of the premetric tensor field are $\mathcal{SL}\left(n+1,\mathbb{R}\right)$
tensor valued 0-forms. The~infinitesimal transformation of the premetric
field under local $\mathcal{SL}\left(n+1,\mathbb{R}\right)$ is given
by 
\begin{equation}
\delta g^{AB}=\mathring{\lambda}_{\,\,\,C}^{(A}g^{CB)},\,\,\,\,\,\,\delta g_{AB}=\mathring{\lambda}_{\,\,(A}^{C}g_{CB)}.
\end{equation}

With the help of the premetric tensor field, we can define~the~following
combined~structure, 
\begin{equation}
\mathring{\mathcal{R}}^{AB}=\mathring{\mathcal{R}}_{\,\,C}^{[A}g^{CB]},
\end{equation}
and one can easily show that the gauge variation of $\mathring{\mathcal{R}}^{AB}$
is given by 
\begin{eqnarray}
\delta\mathring{\mathcal{R}}^{AB} & = & \mathring{\lambda}_{\,\,E}^{[A}\mathring{\mathcal{R}}^{EB]}.
\end{eqnarray}

Now, we can define a shifted curvature as follows 
\begin{equation}
\mathcal{Y}^{AB}=\mathring{\mathcal{R}}^{AB}-\mu\mathcal{F}^{AB},\label{eq: shifted_con_msl5}
\end{equation}
where $\mu$ is a dimensionless arbitrary constant. This object also
has the following gauge transformation, 
\begin{equation}
\delta\mathcal{Y}^{AB}=\mathring{\lambda}_{\,\,E}^{[A}\mathcal{Y}^{EB]}.
\end{equation}

At this point, to~find a gravitational action, similar to the previous
section, we again consider the Stelle--West model. Thus, we write
the following action in $5$ dimensions, 
\begin{equation}
S_{MSL}=\sigma\int V^{E}\eta_{ABCDE}\mathcal{Y}^{AB}\wedge\mathcal{Y}^{CD}+\alpha\left(c^{2}-V_{A}V^{A}\right),\label{eq: action_msl5}
\end{equation}
where, similar to the previous section, $V^{A}$ and $V_{A}=V^{C}g_{CA}$
are 0-form non-dynamical five-vector fields with respect to the special-linear
group in $5$ dimensions and satisfy $\delta V^{A}=\mathring{\lambda}_{\,\,\,C}^{A}V^{C}$
and $\delta V_{A}=\mathring{\lambda}_{\,\,\,A}^{C}V_{C}$, respectively.
Moreover, the transformation of the fully anti-symmetric tensor $\eta_{ABCDE}$
can be given as 
\begin{equation}
\delta\eta_{ABCDE}=-\mathring{\lambda}_{\,\,A}^{F}\eta_{FBCDE}-\mathring{\lambda}_{\,\,B}^{F}\eta_{AFCDE}-\mathring{\lambda}_{\,\,C}^{F}\eta_{ABFDE}-\mathring{\lambda}_{\,\,D}^{F}\eta_{ABCFE}-\mathring{\lambda}_{\,\,E}^{F}\eta_{ABCDF}.
\end{equation}

Here, making use of Equation~(\ref{eq: msl_decompose}), one can
decompose the special linear transformation parameter as $\mathring{\lambda}_{\,\,B}^{A}=\tilde{\lambda}_{\,\,B}^{A}-\frac{1}{n}\delta_{\,\,B}^{A}\tilde{\lambda}$.
Thus, it can be easily shown that $\delta\eta_{ABCDE}=0$.

The action in Equation~(\ref{eq: action_msl5}) is a slightly modified
version of Equation~(\ref{eq: action_1_mgl}) due to the special-linear
group symmetry and its equations of motion being the same as that
of Equation~(\ref{eq: eom_mgl5}).

\subsubsection*{A Gravitational Action in Four~Dimensions}

To find a gravitational model in four dimensions, we will use the
action in Equation~(\ref{eq: action_msl5}). First of all, we assume
that the premetric tensor field $g^{AB}$ is diagonal unless otherwise
indicated. The~special-linear connection tensor then decomposes into
anti-symmetric and symmetric parts via the premetric tensor as 
\begin{equation}
\mathring{\omega}_{\,\,B}^{A}=\omega^{AC}g_{CB}+v^{AC}g_{CB},
\end{equation}
where $\omega^{AC}\left(x\right)$ is anti-symmetric and $v^{AC}\left(x\right)$
is symmetric, with respect to the indices. Then we can decompose the
curvature 2-forms as follows 
\begin{equation}
\mathring{\mathcal{R}}^{AB}=\mathring{R}^{AB}+E^{AB},
\end{equation}
where $R^{AB}$ is the anti-symmetric part and $E^{AB}$ is the symmetric
part and their explicit expressions are given by 
\begin{eqnarray}
\mathring{R}^{AB} & = & d\omega^{AB}+\omega_{\,\,C}^{A}\wedge\omega^{CB}+v_{\,\,C}^{A}\wedge v^{CB},\nonumber \\
E^{AB} & = & dv^{AB}+\omega_{\,\,C}^{(A}\wedge v^{CB)}.
\end{eqnarray}

In this regard, the~Maxwell curvature 2-form can also be written
as 
\begin{equation}
\mathcal{F}^{AB}=dB^{AB}+\omega_{\,\,C}^{[A}\wedge B^{CB]}+v_{\,\,C}^{[A}\wedge B^{CB]},
\end{equation}
and the shifted curvature 2-form takes the following form 
\begin{equation}
\mathcal{Y}^{AB}=\mathring{R}^{AB}+E^{AB}-\mu\mathcal{F}^{AB}.\label{eq: shifted_con_msl5-1}
\end{equation}

{Now}, we will {reduce the space-time dimension from 5 to 4}.
Since $g^{A4}=0$, one can identify the field $g^{ab}\left(x\right)$
as the metric tensor in the four dimensions with the variation $\delta g^{ab}=\mathring{\lambda}_{\,\,\,c}^{(a}g^{cb)}$,
under infinitesimal gauge transformation. Under~these circumstances,
the action in Equation~(\ref{eq: action_msl5}) spontaneously breaks
down to the MacDowell--Mansouri{-}like action~\citep{MacDowell:1977jt}
\begin{eqnarray}
S_{MSL} & = & -\frac{1}{4\kappa\Lambda}\int\eta_{abcd}\mathcal{Y}^{ab}\wedge\mathcal{Y}^{cd}\nonumber \\
 & = & -\frac{1}{4\kappa\Lambda}\int\eta_{abcd}\left(\mathring{\mathcal{R}}^{ab}\wedge\mathring{\mathcal{R}}^{cd}-2\mu\mathring{\mathcal{R}}^{ab}\wedge\mathcal{F}^{cd}+4\mu\mathcal{F}^{ab}\wedge\mathcal{F}^{cd}\right).\label{eq: action_msl4}
\end{eqnarray}
where we again used $V^{A}=\left(c,0,0,0,0\right)$ and $\sigma c=-\frac{1}{4\kappa\Lambda}$.
We also note that $\mathring{\mathcal{R}}^{ab}=\mathring{R}^{ab}+E^{ab}$
is asymmetric under the interchange of indices, so only the anti-symmetric
part of $\mathring{\mathcal{R}}^{ab}$ contributes to the equations
of motion in the presence of the fully anti-symmetric tensor $\eta_{abcd}$.
At this point, one can redefine the shifted curvature in terms of
an anti-symmetric part of $\mathring{\mathcal{R}}^{ab}$ by excluding
the symmetric $E^{ab}$ part as 
\begin{eqnarray}
Y^{ab} & = & \boldsymbol{R}^{ab}-\mu F^{ab},\label{eq: shifted_con_msl4-1}
\end{eqnarray}
where the new objects $\boldsymbol{R}^{ab}$ and $F^{ab}$ are given
by

\begin{eqnarray}
\boldsymbol{R}^{ab} & = & R^{ab}\left(\omega\right)+v_{\,\,c}^{a}\wedge v^{cb},\nonumber \\
F^{ab} & = & dB^{ab}+\omega_{\,\,c}^{[a}\wedge B^{cb]}+v_{\,\,c}^{[a}\wedge B^{cb]}-\lambda e^{a}\wedge e^{b}-\lambda r^{a}\wedge r^{b}+2\lambda b^{a}\wedge b^{b},
\end{eqnarray}
{with} the following linear combinations 
\begin{eqnarray}
e^{a}\left(x\right) & = & \frac{1}{\sqrt{|\Lambda|}}(\omega^{a5}-\mu B^{a5}),\label{eq: e}
\end{eqnarray}
\begin{eqnarray}
r^{a}\left(x\right) & = & \frac{1}{\sqrt{|\Lambda|}}(v^{a5}-\mu B^{a5}),\label{eq: v}
\end{eqnarray}
\begin{equation}
b^{a}\left(x\right)=\frac{\mu}{\sqrt{|\Lambda|}}B^{a5}.\label{eq: b}
\end{equation}
The field $e^{a}\left(x\right)$ may be identified as the generalized
vierbein field and the remaining objects are additional vector fields.
If one assumes $\omega^{ab}\left(x\right)$ to be the Riemannian connection
1-form, then one can identify $R^{ab}\left(\omega\right)=d\omega^{ab}+\omega_{\,\,c}^{a}\wedge\omega^{cb}$
to be {a} Riemann curvature 2-form. Making use of the new definition
of the shifted curvature Equation~(\ref{eq: shifted_con_msl4-1}),
the~action Equation~(\ref{eq: action_msl4}) can be written as follows,
\begin{eqnarray}
S_{MSL} & = & -\frac{1}{2\kappa}\int\eta_{abcd}\left[\frac{1}{2\Lambda}R^{ab}\left(\omega\right)\wedge R^{cd}\left(\omega\right)+\frac{1}{\Lambda}R^{ab}\left(\omega\right)\wedge v_{\,\,e}^{c}\wedge v^{ed}\right]\nonumber \\
 &  & -\eta_{abcd}\left[\frac{\mu}{\Lambda}R^{ab}\left(\omega\right)\wedge F^{cd}+\frac{\mu}{\Lambda}F^{ab}\wedge v_{\,\,e}^{c}\wedge v^{ed}-\frac{\mu^{2}}{2\Lambda}F^{ab}\wedge F^{cd}\right].
\end{eqnarray}
The first term in the first line is the topological Gauss--Bonnet
term with respect to $R^{ab}\left(\omega\right)$, which does not
contribute to the equations of motion. In~the second line, the~first
term contains generalized Einstein--Hilbert action together with
some additional interaction terms. The~third term contains the generalized
cosmological constant term and additional interaction terms between
$B^{ab}\left(x\right)$, $e^{a}\left(x\right)$, $v^{a}\left(x\right)$,
and~$c^{a}\left(x\right)$ fields.

\section{Conclusions\label{sec: 4}}

In this work, we investigated the Maxwell extensions of the general-linear
group $\mathcal{GL}\left(n,\mathbb{R}\right)$ and the special-linear
group $\mathcal{SL}\left(n+1,\mathbb{R}\right)$. Firstly, we presented
the Maxwell extension of the $\mathcal{GL}\left(n,\mathbb{R}\right)$
group and we showed that this extension leads to the Maxwell algebra
\citep{soroka2005tensor,azcarraga2011generalized}, Maxwell--Weyl
algebra \citep{cebeciouglu2014gauge}, and Maxwell--Affine algebra~\citep{cebeciouglu2015maxwell}
when we chose appropriate subalgebras of $\mathcal{GL}\left(5,\mathbb{R}\right)$.
In this context, we also derived a new type of Maxwell--Affine algebra
endowed with a metric tensor in Equation~(\ref{eq: mga4_long}).
Moreover, we constructed the gauge theory of gravity based on the
extended case (\ref{eq: mgl5}) and we wrote down a Stelle--West-like
gauge invariant gravitational action in $5$ dimensions.

Then we analyzed the action in Equation~(\ref{eq: action_SW}) for
two conditions under the dimensional reduction from $5$ to $4$ dimensions
(an alternative approach to the Stelle--West method, one can use
the coset space dimensional reduction method~\citep{Kapetanakis:1992Cosetspace,Manolakos:2019GaugeTheories}).
In the first one, we kept the affine characteristics of the construction
and we found an action in Equation~(\ref{eq: action_mga(4R)}) which
has {a} similar structure as that of the action given in~\citep{cebeciouglu2015maxwell},
but with a more general gravitational action. In~the second condition,
we employed the premetric tensor as a diagonal metric tensor for the
tangent space and found a generalized gravitational theory which includes
the Einstein--Hilbert-like action together with the generalized cosmological
term in Equation~(\ref{eq: Action_CC}).

Secondly, we demonstrated the tensor extension of the $\mathcal{SL}\left(n+1,\mathbb{R}\right)$
group in the context of Maxwell symmetry. By~using suitable subalgebras
of $\mathcal{SL}\left(n+1,\mathbb{R}\right)$, we obtained the Maxwell-special-affine
algebra~\citep{kibarouglu2019maxwellSpecial} and additional new
kinds of extended algebras. The~algebra given in Equation~(\ref{eq: algebra_metalinear})
corresponds to the Maxwell extension of the meta-linear algebra~\citep{Mielke:2011_Spontaneously_SL5R,Mielke:2012_Einstein_Weyl}
and in Equation~(\ref{eq: msa4_long}) we presented a new kind of
Maxwell-special-affine algebra endowed with a metric tensor. Moreover,
we constructed the gauge theory of gravity and we derived a modified
gravitation theory in five dimensions from the Stelle--West-like
action. We then reduced the dimension of the action in Equation~(\ref{eq: action_msl5})
from $5$ to $4$ dimensions, and we derived a gravitational action
which contains the Einstein--Hilbert term, a~generalized cosmological
term together with additional terms. This result generalizes the results
given in \citep{kibarouglu2019maxwellSpecial}.

Finally, we can infer that the Maxwell extension of $\mathcal{GL}\left(n,\mathbb{R}\right)$
and $\mathcal{SL}\left(n+1,\mathbb{R}\right)$ algebras lead to the
derivation of richer gravity theories which may include generalized
cosmological constant terms and some additional terms in $4$-dimensional
space-time. It is well-known that dark energy may be described by
adding the cosmological constant term to the standard Einstein--Hilbert
action, so the gauge theory of the Maxwell extended algebras may play
an important role to explain dark energy phenomenon.
\begin{acknowledgments}
This study is supported by the Scientific and Technological Research
Council of Turkey (TÜB\.{I}TAK) Research Project No: 118F364.
\end{acknowledgments}

\bibliography{Maxwell_GL_SL}

\end{document}